\documentclass[12pt,preprint]{aastex}

\usepackage{lscape}
\input{epsf}

\slugcomment{Accepted for publication in the Astronomical Journal}

\shorttitle{Spectroscopic confirmation of UV-bright white dwarfs}
\shortauthors{L\'epine, Bergeron, \& Lanning}

\begin{document}

\title{Spectroscopic confirmation of UV-bright white dwarfs from the
  Sandage Two-Color Survey of the Galactic Plane}

\author{S\'ebastien L\'epine\altaffilmark{1}, P.
  Bergeron\altaffilmark{2}, and Howard H. Lanning\altaffilmark{3,4}}

\altaffiltext{1}{Department of Astrophysics, Division of Physical
  Sciences, American Museum of Natural History, Central Park West at
  79th Street, New York, NY 10024, USA; lepine@amnh.org}

\altaffiltext{2}{D\'epartement de Physique, Universit\'e de Montr\'eal, C.P.~6128, 
Succursale Centre-Ville, Montr\'eal, QC H3C 3J7, Canada}

\altaffiltext{3}{National Optical Astronomical Observatories, 940
  North Cherry Avenue, Tucson, AZ, USA}

\altaffiltext{4}{Visiting Astronomer, Kitt Peak National Observatory,
  Kitt Peak, AZ, USA}

\begin{abstract}
We present spectroscopic observations confirming the identification of
hot white dwarfs among UV$-$bright sources from the Sandage Two-color
Survey of the Galactic plane and listed in the Lanning (Lan) catalog
of such sources. A subsample of 213 UV bright Lan sources have been
identified as candidate white dwarfs based on the detection
of a significant proper motion. Spectroscopic observations of 46
candidates with the KPNO 2.1m telescope confirm 30 sources to be
hydrogen white dwarfs with subtypes in the DA1$-$DA6 range, and with one
of the stars (Lan 161) having an unresolved M dwarf as a
companion. Five more sources are confirmed to be helium white dwarfs,
with subtypes from DB3 to DB6. One source (Lan 364) is identified as a
DZ 3 white dwarf, with strong lines of calcium. Three more stars are
found to have featureless spectra (to within detection limits), and
are thus classified as DC white dwarfs. In addition, three sources are
found to be hot subdwarfs: Lan 20 and Lan 480 are classified as sdOB,
and Lan 432  is classified sdB.
The remaining four objects are found to be field F star
interlopers. Physical parameters of the DA and DB white dwarfs are
derived from model fits.
\end{abstract}


\section{Introduction}

Areas of the sky at low Galactic latitudes remain relatively
unexplored for white dwarf stars. The Villanova catalog of
spectroscopically identified white dwarfs \citep{MS99}, recently
updated\footnote{http://www.astronomy.villanova.edu/WDCatalog/index.html},
shows a dearth of objects in the Galactic latitude range $-25<b<25$. Yet
white dwarfs should be quite numerous in the plane of the Milky Way,
especially the younger and more massive hot white dwarfs associated
with the young disk population. The 122 very nearby white dwarfs ($d
< 20$ pc) identified by \citet{Ho08} appear to be distributed
uniformly over the sky, which suggests relative completeness; however,
ongoing surveys are turning up additional white dwarfs within this
range \citep{Su07,Le09}, and it is clear that the white dwarf census
is significantly incomplete beyond 20 pc. Many more white dwarfs
remain to be identified, particularly in the low galactic latitude
regions. This is important if one is to obtain a statistically
complete census within a much larger volume and to track the younger
white dwarfs in local zones of recent star formation \citep{deZ99}.

In recent years, white dwarfs have been identified in large numbers
based on color cuts and/or reduced proper motion selection in deep
optical surveys such as in the Sloan Digital Sky Survey
\citep[SDSS;][]{Ki06,Ei06} and the CFHTLS Deep fields \citep{LKKBA} and also
in photographic material for the GSC-II catalog \citep{Ca06}. These
surveys have however been covering only areas of high Galactic
latitude, and have been mostly sensitive to relatively distant, old
white dwarfs of the thick disk and halo \citep{Ga04,Vi07}. 

Proper motion surveys remain the principal source of discovery for
very nearby white dwarfs. Follow-up studies of older proper motion
catalogs such as the Luyten half-second (LHS) and NLTT
\citep{Lu79a,Lu79b} continue to turn up new white dwarfs
\citep{VK03,RG05}, while new proper motion surveys are filling the
gaps \citep{Ru01,Le03,Sc04,Ro08}. However, because high proper motion
catalogs are more sensitive to nearby high velocity stars, many of the
new white dwarfs are found to be associated with
the Galactic halo \citep{Sc05,Le05}. White dwarfs from the young disk
population are not expected to have extreme proper motions beyond 20
pc, but should be found among the thousands to millions of stars with
small to moderate proper motions. White dwarfs can be identified in
large catalogs of proper motion sources with the use of a reduced
proper motion diagram, in which they populate a distinct locus. The
identification is particularly straightforward when one uses an
optical-to-infrared reduced proper motion diagrams \citep{LS05}, where
the white dwarfs are segregated from field dwarfs and
subdwarfs. Unfortunately, white dwarfs fade out from all-sky infrared
surveys (Two Micron All Sky Survey, DENIS) beyond a short distance
range due to their low intrinsic luminosity in infrared bands.

One promising avenue is the combination of proper motion data with
catalogs of UV-bright sources. Faint UV-bright sources, in particular,
are unlikely to have significant proper motions unless they are white
dwarfs. We have recently compiled proper motion data for UV-bright
sources identified in a $\sim$2000 deg$^2$ area of the Sandage
two-color survey of the Galactic plane, in an attempt to identify
white dwarf stars at low galactic latitudes. Our analysis indicates
that a reduced proper motion diagram readily identifies candidate
white dwarfs \citep{LL06}.

In this paper, we present the results of our spectroscopic follow-up
survey which confirms the white dwarf status of our UV and proper
motion selected sources. Formal spectral classification is presented
for 46 UV-bright sources, of which 39 are confirmed to be white
dwarfs, three are hot subdwarfs, and the remaining four and field F
dwarf interlopers. Target selection is described in Section 2,
observations are presented in Section 3, results are analyzed in
Section 4, and summarized in the conclusion (Section 5). 

\begin{figure*}
\epsscale{1.1}
\plotone{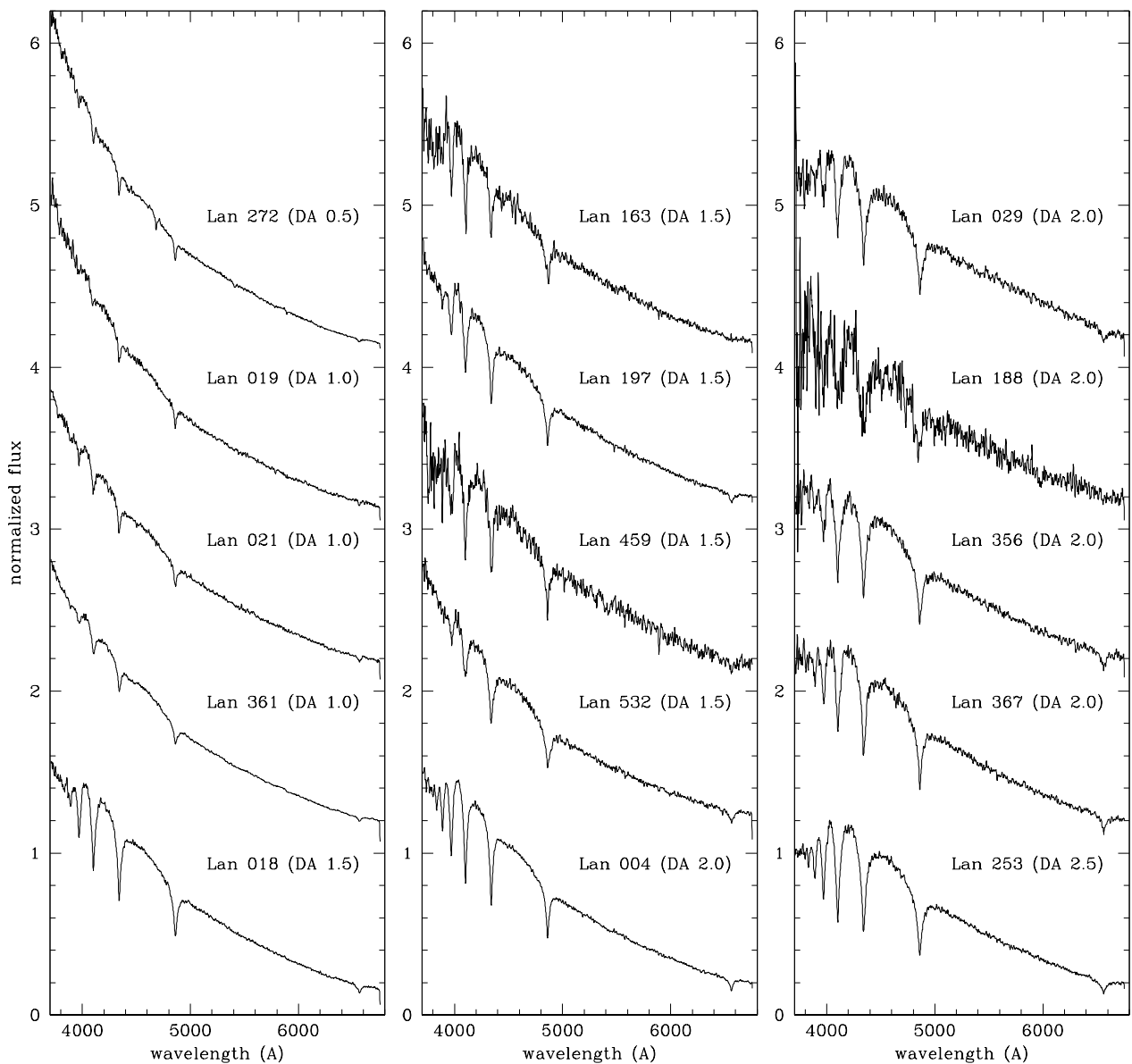}
\caption{Lanning sources classified as DA white dwarfs, compact
  degenerates with hydrogen-line spectra. The spectrum of Lan 272
  also exhibits the He {\sc ii} $\lambda$4686 absorption feature and is
  thus classified as DAO.}
\end{figure*}

\begin{figure*}
\plotone{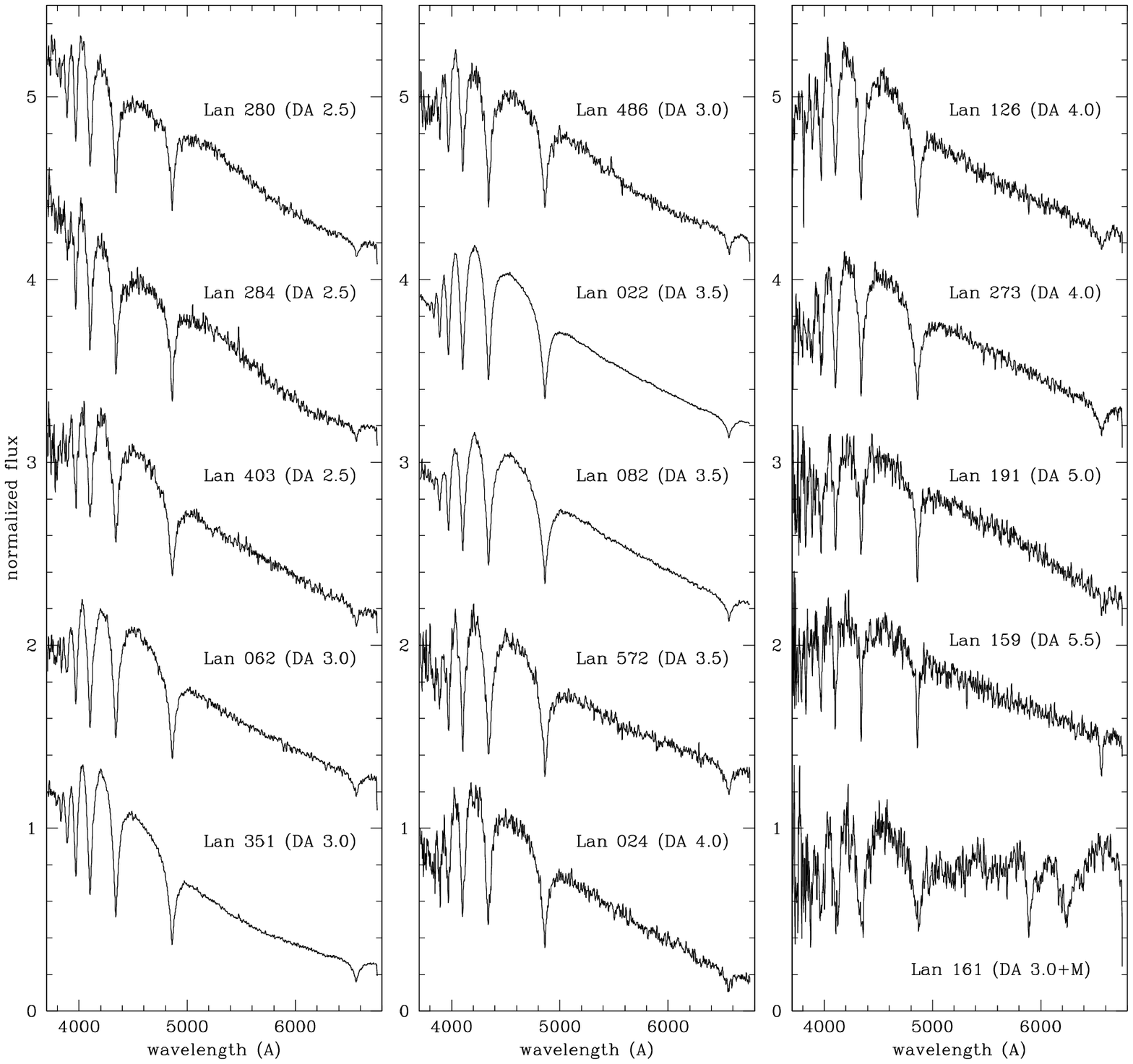}
\caption{Lanning sources classified as DA white dwarfs, compact
  degenerates with hydrogen-line spectra. The star Lan 161 is
  revealed to have an unresolved M dwarf companion.}
\end{figure*}

\section{Target Selection}

The Lanning catalog of UV$-$bright sources compiles objects with very
blue $U-B$ colors, identified in the Sandage two-color survey of the
Galactic plane. UV$-$bright sources were identified by visual scanning,
in 48 of the 124 two-color photographic plates. With each plate
covering a field of about 43 deg$^2$, the survey extends to
about 2000 deg$^2$ of low Galactic latitude fields. The survey
has produced 734 published sources, ranging in color from $U-B=0$ to
$U-B=-1.5$ (Lanning 1973, Lanning \& Meakes 1994, 1995, 1998, 2000,
2001, 2004). The catalog includes seven identified cataclysmic
variables, along with many verified and probable hot stars such as
white dwarfs, Be stars, hot subdwarfs, and central stars of planetary
nebulae. However, only a few of the published Lanning stars have
formal spectral classification. Counterparts to Lanning stars were
identified in the USNO-B1 catalog of Monet et al. (2003) with 213
sources found to have proper motions larger than 10 mas yr$^{-1}$;
these have been suggested to be mostly white dwarfs within
$\approx200$ pc of the Sun \citep{LL06}. These stars were targeted for
follow-up spectroscopy.

\section{Observations}

Observations were performed on the nights of 2007 October 15$-$18,
from the 2.1 m telescope on Kitt Peak. In four nights, a total of 46
target stars were observed with the GoldCam CCD spectrograph. We used
the ``240'' grism, blazed at 5500~\AA, to obtain spectra with a
resolution of 1.5~\AA\ pixel$^{-1}$. Calibration arcs of HeNeAr were
observed for wavelength calibration. Four standard stars were also
observed for flux calibration.

Reduction was conducted with IRAF, including standard bias
correction, flat fielding, subtraction of the sky background, aperture
extraction, wavelength calibration, and normalization. The resulting
medium-resolution spectra are calibrated to a wavelength accuracy of
0.3~\AA.

Spectra were first examined visually and an initial classification
was made based on comparison with published, spectroscopic sequences
of main-sequence stars \citep{SC92} and hot subdwarfs \citep{Mo90},
and with the spectroscopic atlas of white dwarfs compiled by
\citet{We93}. Most sources were classified as white dwarfs of spectral
class DA (hydrogen-line), DB (helium-line), and DC (featureless). Four
sources were found to be consistent with main-sequence F stars; five
others were found to be consistent with hot subdwarfs
(sdB/sdO). Spectral subtypes for the DA and DB white dwarfs were
assigned on the basis of the temperatures determined by the
atmospheric model fits (see Section 3). Subtypes for the DC white
dwarfs were estimated by comparing the overall spectral energy
distribution with a blackbody profile. Subtypes were assigned
following the general prescription of $50,400/T_{\rm eff}$, with the
resulting value rounded up to the nearest half subtype
\citep{We93}. Table 1 lists all 46 sources observed, their positions
and proper motions, their magnitudes on the photographic POSS-II
plates as obtained from the USNO-B1.0 catalog \citep{Monet2003}, and
their assigned spectral type.

\begin{figure*}
\plotone{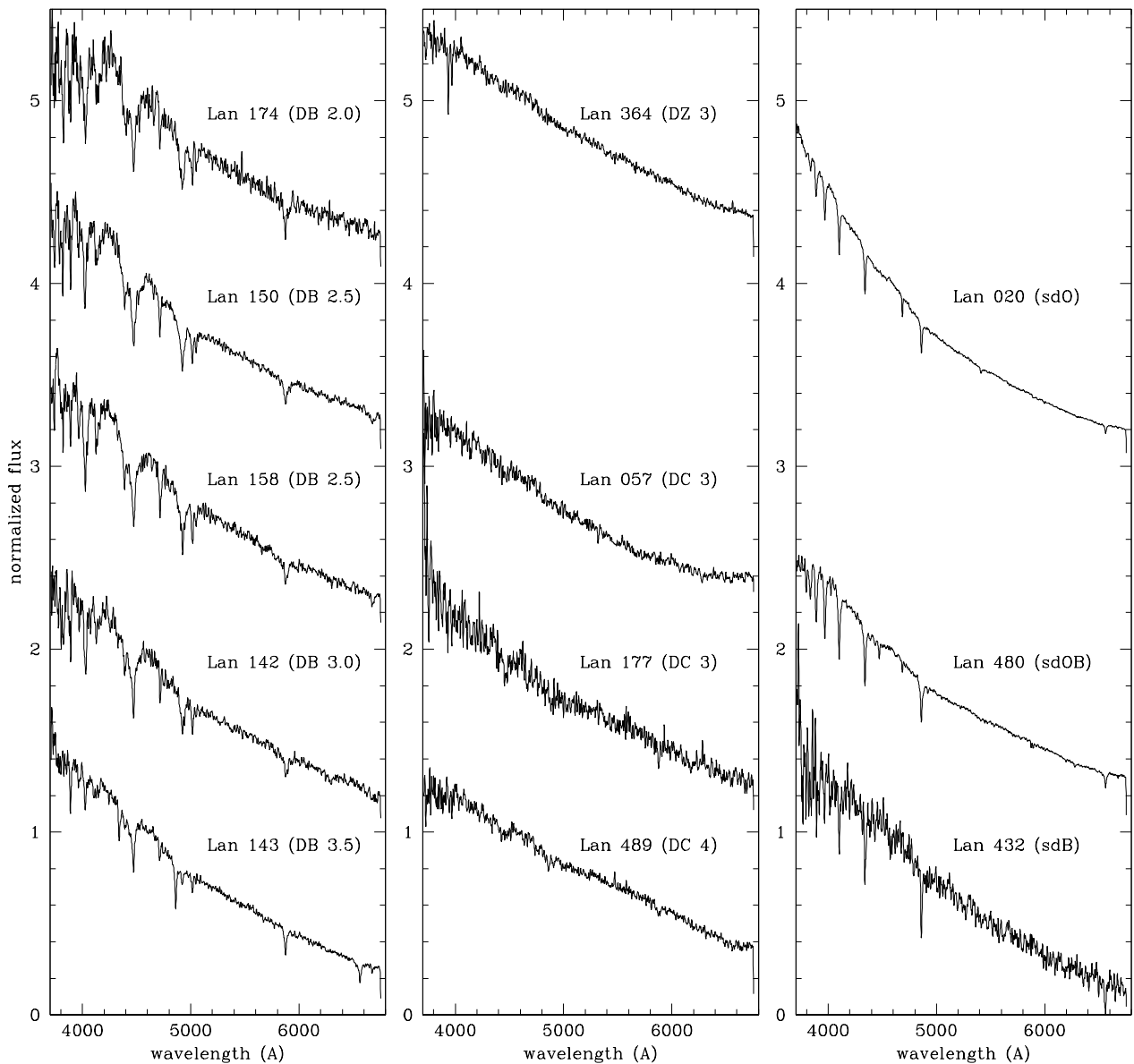}
\caption{Lanning sources classified as DB, DZ, and DC white
  dwarfs. The DB white dwarfs have spectra dominated by neutral helium
  lines, while the DZ shows lines of calcium. The DC white dwarfs
  appear featureless to within our detection limits. Also shown are
  the stars we classify as hot subdwarfs (sdO and sdB), whose
  gravities are lower than main-sequence stars but not as extreme as
  in the white dwarfs.}
\end{figure*}

\begin{figure}
\plotone{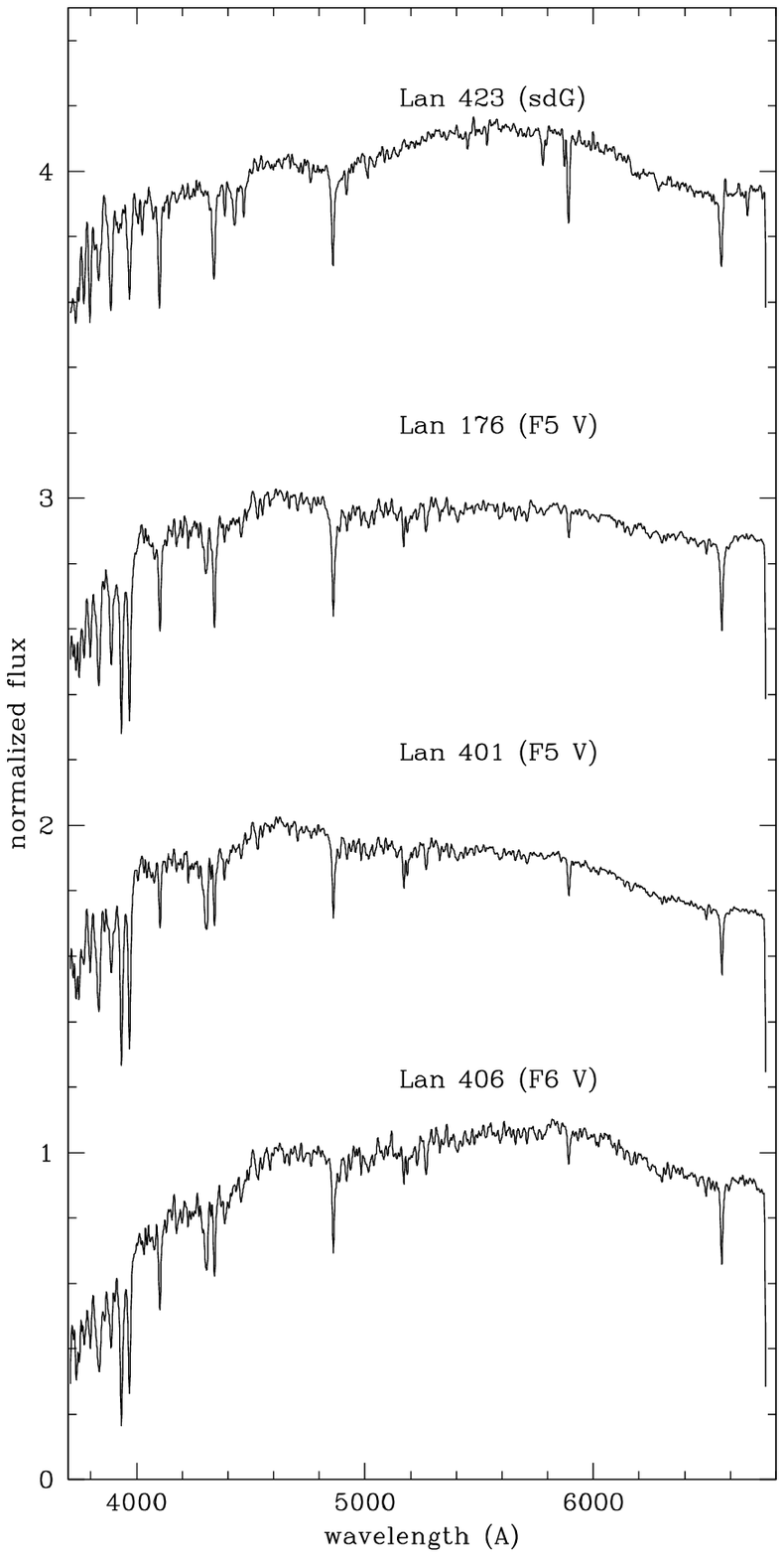}
\caption{Lanning sources initially flagged as probable white dwarfs,
  but revealed to be field F stars. One of the sources (Lan 423) has
  the Na $\lambda$ line as strong as H$\alpha$, which would
  normally indicate a spectral type G, but lines of Mg $\lambda$ are
  absent of weak. We interpret this as the star being metal poor, and
  classify it as a G subdwarf (sdG).}
\end{figure}

\section{Comparison to previously known objects}

A search of the literature found published, formal spectral subtypes
for five of the stars. The subtypes are listed in Table 1 along with
the bibliographic reference. The previously reported subtypes are
broadly consistent with our own classification. 

These stars include Lan 18, a fairly well-known DA star first
classified as DA2 by \citet{LS77}; also Lan 020, which was first
identified as an sdO by \citet{Walker1981}. In addition, six of our
sources had also been previously observed by \citet{Era2002}, who
provided a broad ``spectral class'' diagnostic to indicate whether the
star appeared to be a hot object or not. Both Lan 19 and Lan 21 were
noted to be ``Hot, high gravity?,'' Lan 22 was classified as
``O7-O9,'' Lan 367 was classified as ``B0-B3,'' Lan 423 was noted to
be ``Early B,'' Lan 401 was identified as a ``late F-G,'' and Lan 406
as a ``late-F,'' while Lan 423 was identified as ``early-B'' and Lan
459 was simply classified as ``DA.'' Those impressions are also noted
in Table 1; they are generally consistent with our own spectral types.

It is also interesting that Lan 021 is associated to a possible
planetary nebula by \citet{Kron2006}. Likewise, Lan 272 was classified
as a possible central star of planetary nebula by \citet{Chromey1979}.

\section{Atmospheric model fits}

Our model atmospheres and synthetic spectra for DA stars are described
at length in \citet{tremblay10} and references therein; these make
use, in particular, of the improved Stark profiles for the hydrogen
lines developed by \citet{tb09}.  Non-local thermodynamic equilibrium
(NLTE) effects are explicitly taken into account above $T_{\rm
  eff}=20,000$~K and energy transport by convection is included in
cooler models following the ML2/$\alpha=0.8$ prescription of the
mixing-length theory. Our model atmospheres and synthetic spectra for
DB stars are similar to those described in \citet{beauchamp96}, which
include the improved Stark profiles of neutral helium of
\citet{beauchamp97}.  Our fitting technique relies on the nonlinear
least-squares method of Levenberg$-$Marquardt, which is based on a
steepest descent method. The model spectra (convolved with a Gaussian
instrumental profile) and the optical spectrum of each star are first
normalized to a continuum set to unity. The calculation of $\chi ^2$
is then carried out in terms of these normalized line profiles
only. Atmospheric parameters \--- $T_{\rm eff}$, $\log g$, as well as
${\rm H}/{\rm He}$ for DBA stars or ${\rm He}/{\rm H}$ for
DAO stars \--- are considered free parameters in the fitting procedure.


\section{Classification and spectroscopic analysis}

Spectra of the sources classified as DA (hydrogen-line)
white dwarfs are displayed in Figures 1 and 2. The 28 stars
show the characteristic, broad lines of the hydrogen Balmer series;
Lan 272 also exhibits the He {\sc ii} $\lambda$4686 absorption feature
and is thus classified as a DAO star. Elements of the atmospheric
model fits are reported in Table 2. The spectral subtypes
listed in Table 1 are based on the model-determined effective
temperatures listed here. Additionally, the star Lan 161 is revealed to
be a binary star, consisting of a white dwarf with a low-mass, main
sequence companion; the star displays the characteristic signature of
broad Balmer lines on the blue side, but redward of 5000 \AA\ the
spectrum clearly displays the broad molecular features usually observed
in M dwarfs \citep{KHM91,RHG95,Le03}, in particular a deep CaOH band
around 6250 \AA.

Spectra of the five objects classified as DB (helium-line) white
dwarfs are displayed in Figure 3; the spectrum of Lan 143 also
shows hydrogen lines and is thus classified DBA. Elements of the
atmospheric fits are provided in Table 3 and include estimates of the
hydrogen abundances, or upper limits, based on the presence or absence
of H$\alpha$.

Figure 3 also shows the stars with featureless spectra, which we
identify as DC white dwarfs, along with the spectrum of Lan 364, which we
classify as a DZ (white dwarf with traces of metal) based on detection
of a Ca {\sc ii} $\lambda\lambda3933-3968$ doublet. These stars are not
suited for atmospheric model fits without accurate color information,
and their physical parameters are left undetermined.

The three stars identified as hot subdwarfs also have their spectra
displayed in Figure 3. All show relatively narrower lines of the hydrogen
Balmer series indicative of a lower surface gravity. Their gravities
are however significantly higher than in main-sequence stars.

Finally, spectra of the four objects classified as main
sequence F stars are shown in Figure 4. The star Lan 423 has a spectrum
consistent with a main-sequence F star, but lacks strong absorption in
the extreme blue, which is interpreted to be the possible signature of
a metal-poor atmosphere.

Atmospheric model fits yield estimates of the ages and masses of the
DA and DB white dwarfs. Hot white dwarfs tend to be relatively young,
and this is confirmed by the model fits. Figure 5 shows the location
of the stars in the mass-effective temperature diagram, and compares
it to theoretical isochrones. For DA stars with $T_{\rm eff} > 30,000$~K, we
use the carbon-core cooling models of \citet{wood1995} with thick
hydrogen layers of $q({\rm H})\equiv M_{\rm H}/M_{\star}=10^{-4}$ and
$q({\rm He})=10^{-2}$, while for $T_{\rm eff} < 30,000$~K, we use cooling models
similar to those described in \citet{fon01} but with carbon$-$oxygen
cores. For DB stars, we rely on similar models but with thin hydrogen
layers of $q({\rm H})=10^{-10}$ representative of helium-atmosphere
white dwarfs. The distribution shows that most of our stars have been
in the white dwarf phase for less than 1 Gyr. Model-fits yield
relatively low-masses for the youngest white dwarfs, some apparently
below 0.4 $M_{\sun}$, and these are most likely unresolved double
degenerate binaries that are the result of common envelope evolution.

\begin{figure}
\epsscale{0.8}
\plotone{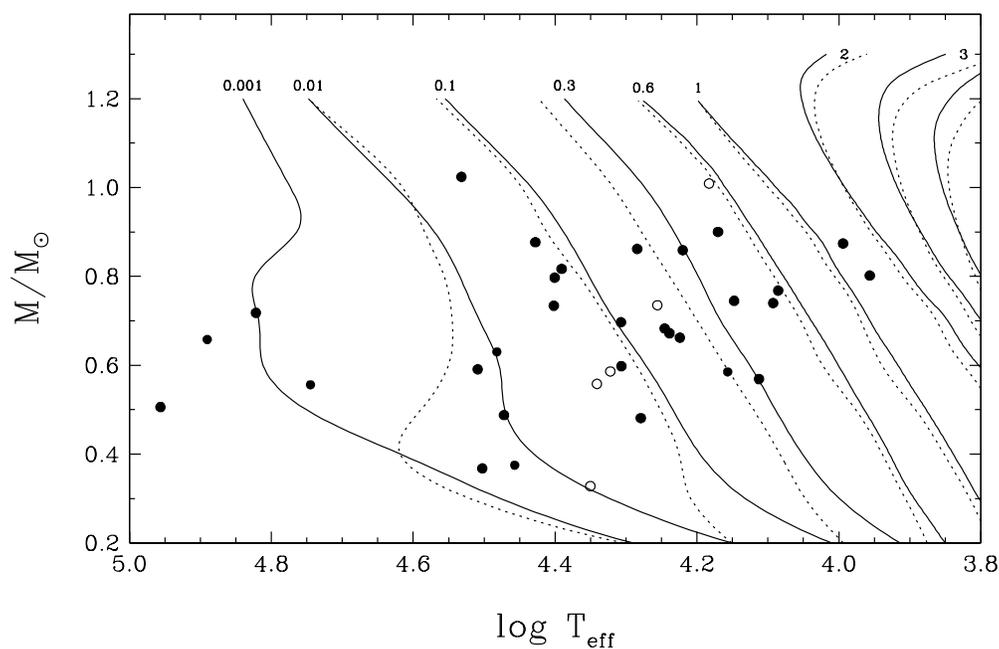}
\caption{Isochrones of the white dwarf evolutionary models in the
  mass/effective temperature diagram. The isochrones are labeled in
  billions of years (Gyr); DA white dwarf isochrones are plotted as
  continuous lines, while the slightly different DB isochrones are
  plotted as dotted lines. Estimates for the white dwarfs presented in
  this paper are noted as filled circles for the DA white dwarfs, and
  as open circles for the DBs. Most of our stars are found to have
  been in the white dwarf phase for less than 1 billion years.}
\end{figure}


\section{Kinematic analysis}

By combining the measured proper motions and the distances
estimated using the atmospheric model fits, we calculate transverse
velocities for all the DA and DB white dwarfs. These velocities are
two-dimensional vectors, which are projections of the
three-dimensional motion
vectors of the stars in the plane of the sky. All of our stars are
located at low Galactic latitudes ($|b|<$10.0) which means that the
proper motion in the direction of the galactic coordinate $b$ is
effectively equivalent to the component of motion $W$ (the
component of velocity normal to the plane of the Galaxy.) The
component of motion in the direction of $l$ is then a combination of
the components of motion $U$ (in the direction of the Galactic center)
and $V$ (in the direction of Galactic rotation). 

Generally, the components of $U$, $V$, and $W$ can be calculated using
\begin{equation}
U = 4.74 d ( \mu_l \sin(l) - \mu_b \cos(l) \sin(b) ) + V_r \cos(l) \cos(b) 
\end{equation}
\begin{equation}
V = 4.74 d ( \mu_l \cos(l) - \mu_b \sin(l) \sin(b) ) + V_r \sin(l) \cos(b)
\end{equation}
\begin{equation}
W = 4.74 d \mu_b \cos(b) + V_r \sin(b).
\end{equation}

For the case where radial velocities are unknown, it is possible
to estimate with reasonably good accuracy the components of motion for
any two pairs ($UV$, $UW$, or $VW$) by selecting subsamples of stars at
particular locations on the sky. For example, stars with
$|b|\approx90.0$ have $\cos(b)\approx0$ and from Equations (1) and (2)
we see that the component of radial velocity becomes negligible in the
calculation of both $U$ and $V$. The calculated $U$ and $V$ components of
stars near the Galactic pole will thus be reasonably accurate, even
though the radial velocity component is not known a priori. Likewise,
components of $U$ and $W$ will be a reasonably good representation for
proper motion stars near $l=90,b=0$ or $l=270,b=0$, and components of
$V$ and $W$ will be a reasonable good representation for stars near
$l=0,b=0$ or $l=180,b=0$.

We thus assume all radial velocities to be zero (0.0) and estimate
the components of motion in $U$, $V$, and $W$ using Equations
(1)$-$(3) for all the DA and DB white dwarfs in our sample. We then
separate our objects into two groups: (1) stars with $45<l<135$ and
$225<l<315$, and (2) stars with $135<l<225$ and $315<l<45$. The first
group yields a reasonably accurate distribution in the projected $UW$
velocity plane, while the second groups yield a reasonably good
distribution in the $VW$ plane.

We further separate the samples according to the ages derived from the
model fits. Separate plots are generated for stars with estimated ages
(of the white dwarf phase) of $\tau<$10$^{7.5}$ yr, and stars with
ages $\tau>$10$^{7.5}$ yr. The separation limit corresponds to an age
of $\approx$31 Myr, and essentially sets apart the very recent white
dwarfs from the rest. The velocity-space distributions are shown in
Figure 6. As in Figure 5, the DA white dwarfs are represented with
filled symbols, while the DB white dwarfs are shown as open
symbols. The top panels show the kinematics of the very recent white
dwarfs, while the bottom panels show the kinematics of the older
objects. Transverse velocities are all under 70 km s$^{-1}$, which is
largely consistent with thin disk kinematics. 

\begin{figure}
\plotone{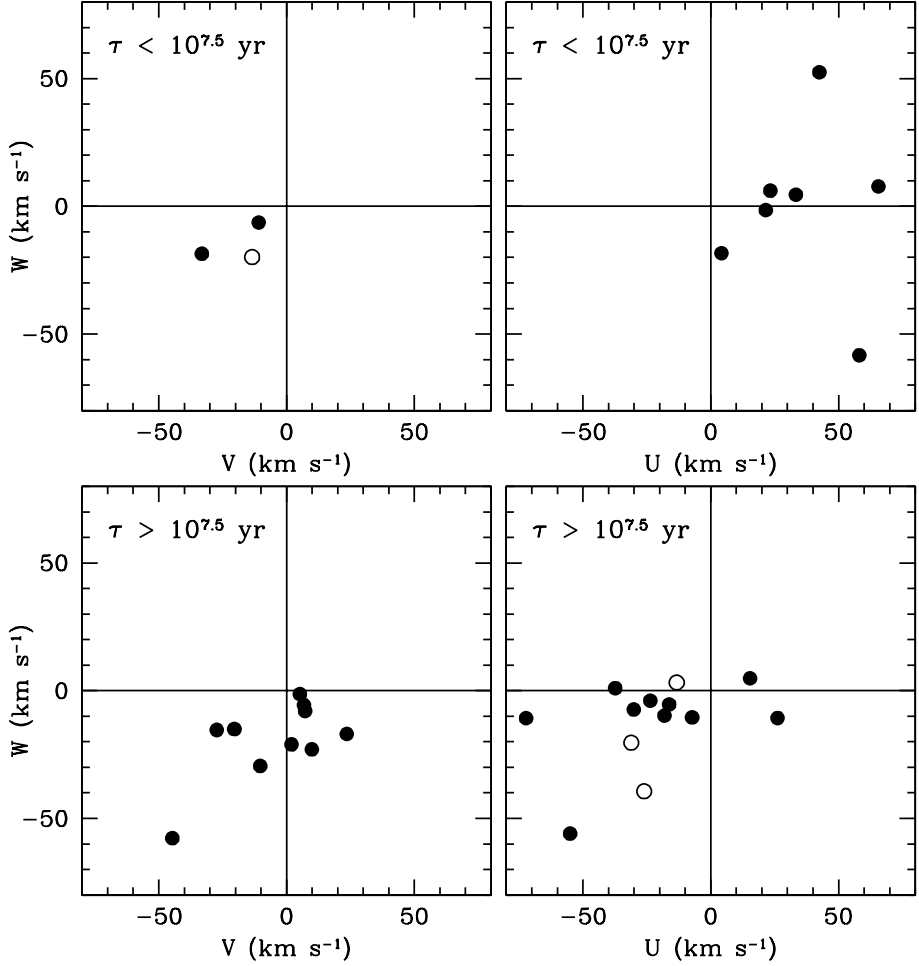}
\caption{Kinematics of the DA and DB white dwarfs (filled and open
  circles, respectively). The left panels show the velocities in the
  direction of Galactic rotation ($V$) and normal to the plane
  ($W$) for stars located in the direction of $l\approx0$ and
  $l\approx180$. Right panels show velocities toward the Galactic
  center ($U$) and normal to the Galactic plane ($W$) for stars located in
  the direction of $l\approx90$ and $l\approx270$. The youngest white
  dwarfs are plotted in the top panels, older ones in the bottom
  panels. Young white dwarfs are all found to have $U>0$ km s$^{-1}$
  while older white dwarfs show the exact opposite trend.}
\end{figure}

An intriguing result is that the youngest white dwarfs appear to
stand out in velocity space: they all show $U>0$ km s$^{-1}$,
whereas the older objects tend to have $U<0$ km s$^{-1}$. The trend
itself has more to do with the orientation of the proper motion
vector, and does not depend much on the accuracy of the distance
estimates. A position value of $U$ signifies that the object is
currently moving toward the Galactic center, while a negative value
means it is moving out. The bulk of the stars in the vicinity of the
Sun have $U<0$ km s$^{-1}$ \citep{Nordstrom2004} including most of the
young moving groups \citep{Bovy2009}. Our results would suggest that
our youngest white dwarfs are not associated with the recent star
formation in the immediate solar neighborhood. Nearby white dwarfs do
appear to have a more symmetric distribution around $U=0$
\citep{KV2006}, which is not so much at odds with our $U>0$ trend
given the small size of our sample. The trend would definitely need to
be confirmed with additional data points. It could be that the very
young white dwarfs in our sample are also the most luminous and the
most distant, with $d\gtrsim200$pc. A possible $U>0$ trend might arise
if these stars are members of more distant young moving groups, with
kinematics very distinct from the local ones. A confirmation of this
trend would require spectrosopic follow-up of a much larger sample of
hot white dwarfs, or a search for kinematic groups beyond the current
distance limit of the {\it Hipparcos} catalog (from which most of the local
moving groups have been discovered).

\begin{figure}
\plotone{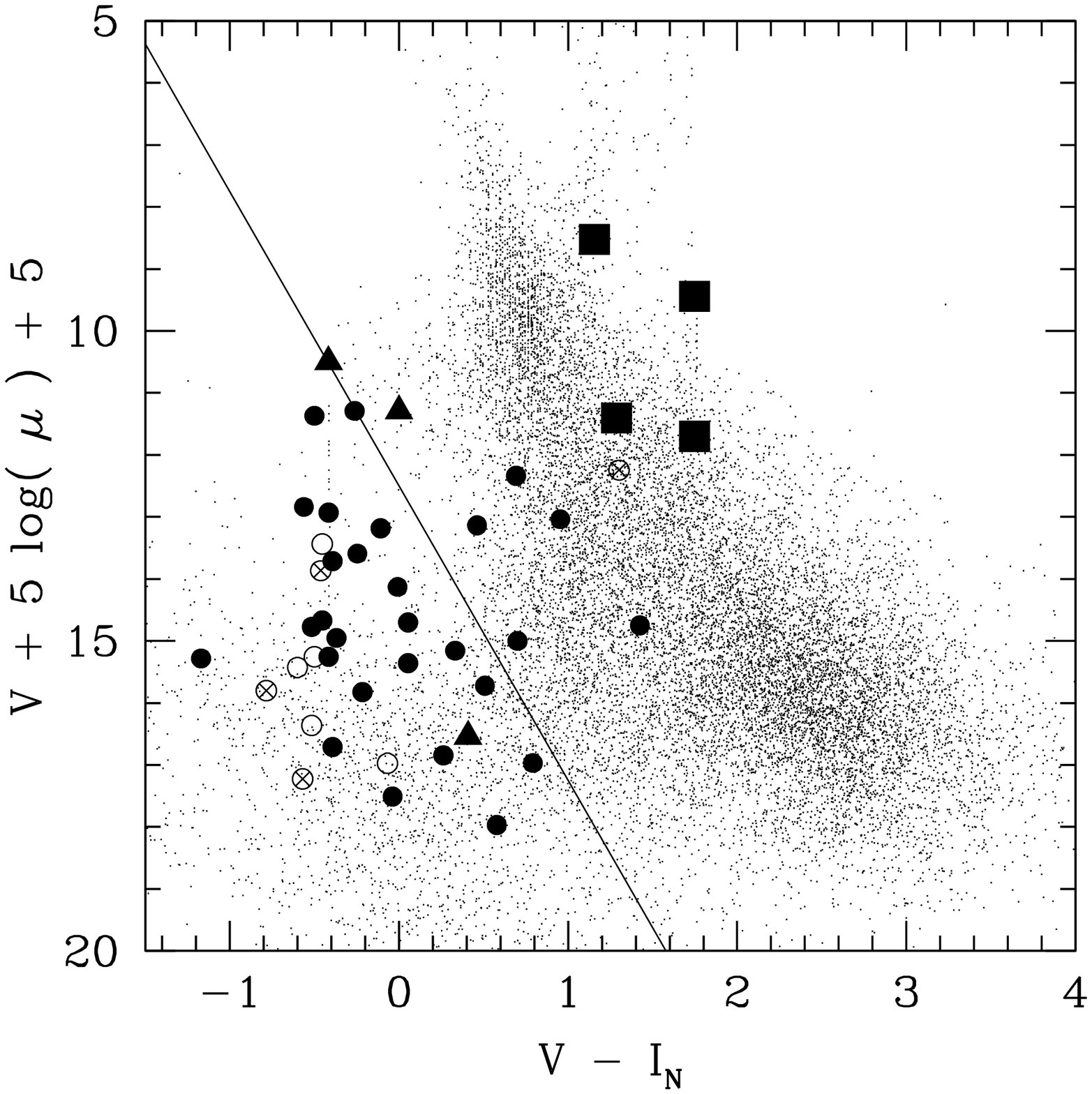}
\caption{Optical-to-infrared reduced proper motion diagram of the
  sources observed in this paper (large symbols). A representative
  sample of stars with large proper motions ($\mu>$150 $\arcsec$
  yr$^{-1}$) is shown for comparison (dots). White dwarfs are shown as
  circles, with filled circles representing the DA, open circles the
  DB, and crossed circles the DZ and DC. Hot subdwarfs (sdO/sdB) are
  plotted as triangles, and main-sequence stars as squares. The
  straight line shows a typical limit used to separate out white dwarf
  (lower left) from main-sequence stars of the disk and halo (upper
  right). Most, but not all, of our confirmed white dwarfs fall on the
  expected side of the line; the hot subdwarfs are blended in 
  with the white dwarfs and cannot be readily identified in the
  diagram.}
\end{figure}

Figure 7 shows the reduced proper motion diagram of the stars observed in
this paper. Different symbols are used to distinguish stars of
different gravity classes (circles for white dwarf, triangles for
subdwarf, squares for main-sequence stars). A representative sample of
15,000 field stars with proper $\mu>20$ mas yr$^{^-1}$, drawn from the
USNO-B1.0 catalog \citep{Monet2003}, is shown for comparison (dots). The
reduced proper motion diagram segregates the main-sequence
stars from the white dwarfs and subdwarfs. In the reduced proper
motion diagram, white dwarfs occupy the lower left of the stellar
locus (blue stars with large reduced proper motion). The straight line
in Figure 7 shows the approximate border separating the white dwarfs from
the main-sequence stars, based on the observed distribution of field
stars. One sees that while the majority of the white dwarfs observed
in our study do indeed fall blueward of the line, some fall to the
red, and would not readily be suspected to be white dwarfs based on
their proper motion and optical colors alone. We suggest that the
combination of a demonstrated UV excess in conjunction with a high
proper motion would be more efficient in identifying hot white dwarfs
at low Galactic latitudes. A reduced proper motion diagram based on
ultraviolet to optical colors may turn out to be the best tool.

One caveat of this method is that interstellar absorption will
significantly reduce the UV brightness of stars beyond a few hundred
parsecs. Selection of hot white dwarfs based on UV excess will thus
work best for relatively nearby objects. The exact range over which
this method will be efficient and/or reliable remains to be
determined.

\section{Conclusions}

Our spectroscopy confirms that, as suggested in \citet{LL06}, most of
the UV-bright stars from the Lanning catalog that also have large
proper motions ($>$10 mas yr$^{-1}$) are for the most part white
dwarfs. Several more Lanning stars remain with no formal spectral
classification. We predict that all the stars in the lower left region
of the reduced proper motion diagram will be revealed as white dwarfs
or hot subdwarfs in future spectroscopic follow-up surveys.

The identification of white dwarfs based on a combination of UV excess
and proper motion holds the potential to significantly increase the
census of young nearby white dwarfs. The strong predictive power
for white dwarfs selected by UV excess suggests that a combination of
proper motion data and broadband UV magnitudes as provided, e.g., by
the {\it Galaxy Evolution Explorer (GALEX)} mission, such as the
recent work by \citet{Ve2011}, would be extremely efficient in
expanding the census of nearby white dwarfs.

However, we note that the {\it GALEX} mission generally avoids fields
at low Galactic latitudes ($|b|<20.0$), which means that these regions
will remain relatively unexplored, until UV imaging will be performed
to cover those potentially rich hot white dwarf hunting grounds.

\acknowledgments

{\bf Acknowledgments}

Howard H. Lanning sadly passed away in 2007 December, just a few
months after carrying out the spectroscopic observations presented in
this paper. He will be missed by his friends and colleagues.

We thank A.~Gianninas for his fits to the two hottest
white dwarfs in our sample and P.~Brassard for providing us with his
grid of subdwarf model spectra. S.L. was supported in this research by
NSF grants AST-0607757 and AST-0908406 at the American Museum of
Natural History.  This work was also supported in part by the NSERC
Canada and by the Fund FQRNT (Qu\'ebec). P.B. is a Cottrell Scholar of
Research Corporation for Science Advancement.


\begin{landscape}
\begin{deluxetable}{lcccccccccll}
\tabletypesize{\scriptsize}
\tablecolumns{12} 
\tablewidth{510pt} 
\tablecaption{Spectroscopically Classified UV-bright Sources} 
\tablehead{ 
\colhead{Name} & 
\colhead{$\alpha$} & 
\colhead{$\delta$} & 
\colhead{$\mu_{\alpha}$} & 
\colhead{$\mu_{\delta}$} & 
\colhead{$m_{B}$} & 
\colhead{$U-B$} & 
\colhead{$B_J$} & 
\colhead{$R_F$} & 
\colhead{$I_N$} & 
\colhead{Sp. Type} & 
\colhead{Sp. Type} 
\\
\colhead{} & 
\colhead{(J2000)} & 
\colhead{(J2000)} & 
\colhead{(mas yr$^{-1}$)} & 
\colhead{(mas yr$^{-1}$)} & 
\colhead{(mag)} & 
\colhead{(mag)} & 
\colhead{(mag)} & 
\colhead{(mag)} & 
\colhead{(mag)} & 
\colhead{(Old)} &
\colhead{(New)}
}
\startdata 
Lan 004 & 01 26 08.8 & +69 01 55.3 & $-$22 &   8  & 16  & $-$0.9 & 15.9 & 16.3 & 16.5 &     & DA 2.0\\
Lan 018 & 18 47 39.1 & +01 57 39.1 &  $-$2 & $-$74  & 12  & $-$0.9 & 12.9 & 13.1 & 12.3 & DA2\tablenotemark{a,b} & DA 1.5\\
Lan 019 & 19 33 49.9 & +18 52 03.1 &  10 & $-$10  & 15.5& $-$0.9 & 15.4 & 15.7 & 15.8 & Hot, high gravity?\tablenotemark{c}    & DA 1.0\\
Lan 020 & 19 43 31.2 & +18 24 35.2 & 3.3 & $-$43.8&     &$<-$0.8 & 12.1 & 12.5 & 12.7 & sdO\tablenotemark{d} & sdO   \\
 Lan 021 & 19 44 59.1 & +22 45 49.2 &  $-$8 & $-$16  & 16  & $-$0.9 & 15.4 & 16.2 & 17.4 & Hot, high gravity?\tablenotemark{c}    & DA 1.0\\
Lan 022 & 20 00 21.6 & +19 07 41.1 &   0 &   0  & 14.7&        & 15.2 & 14.8 & 14.9 & O7-O9\tablenotemark{c}   & DA 3.5\\
Lan 024 & 01 24 05.1 & +69 13 20.3 &  94 & $-$10  & 19  & $-$0.6 & 17.0 & 17.2 & 16.3 &     & DA 4.0\\
Lan 029 & 03 17 51.6 & +53 04 20.0 & $-$14 & $-$16  & 17  & $-$0.6 & 17.4 & 17.6 & 17.5 &     & DA 2.0\\
Lan 057 & 03 10 03.2 & +53 09 24.5 &  28 & $-$70  & 17  & $-$0.3 & 16.6 & 16.2 & 17.2 &     & DC 4  \\
Lan 062 & 03 31 18.1 & +53 03 52.0 & $-$18 &  $-$2  & 17  & $-$0.3 & 16.8 & 17.0 & 17.0 &     & DA 3.0\\
Lan 082 & 21 26 24.9 & +55 13 29.3 & 268 & 192  & 14  & $-$0.3 & 15.7 & 15.0 & 14.8 & DA4\tablenotemark{a} & DA 3.5\\
Lan 126 & 23 39 10.5 & +66 51 45.6 &  44 &  $-$8  & 19.5& $-$0.8 & 16.8 & 17.3 & 18.2 &     & DA 4.0\\
Lan 142 & 02 28 50.0 & +68 35 37.9 &  14 & $-$50 & 17.2 & $-$0.7 & 17.6 & 18.0 & 18.3 &     & DB 3.0\\
Lan 143 & 03 03 27.8 & +68 29 54.1 &  50 & $-$18 & 16.5 & $-$0.3 & 16.8 & 16.8 & 17.4 &     & DBA 3.5\\
Lan 150 & 02 15 34.6 & +64 53 22.2 &  44 & $-$68 & 16.0 & $-$0.5 & 17.8 & 17.0 & 17.5 &     & DB 2.5\\
Lan 158 & 02 19 06.3 & +70 08 39.8 &  36 & $-$40 & 16.5 & $-$1.0 & 16.6 & 16.6 & 17.1 &     & DB 2.5\\
Lan 159 & 01 55 10.9 & +69 42 41.3 &$-$112 &  $-$6 & 17.0 & $-$0.3 & 17.4 & 17.1 & 17.3 &     & DA 5.5\\
Lan 161 & 01 19 41.4 & +68 51 10.8 & $-$26 &  10 & 18.0 & $-$0.5 & 17.8 & 17.2 & 16.1 &     & DA 3.0\\
Lan 163 & 01 33 38.3 & +68 03 32.7 & $-$34 & $-$18 & 16.5 & $-$0.5 & 17.6 & 16.8 & 16.9 &     & DA 1.5\\
Lan 174 & 03 41 17.1 & +62 42 03.1 &   2 & $-$18 & 16.8 & $-$0.7 & 17.1 & 17.2 & 17.6 &     & DB 2.0\\
Lan 176 & 03 56 51.0 & +62 26 20.0 &  10 & $-$18 & 12.5 & $-$0.2 & 12.6 & 11.2 & 10.8 &     & F5 V  \\
Lan 177 & 03 51 15.7 & +61 52 46.4 &  84 & $-$70 & 17.0 & $-$0.5 & 16.8 & 17.3 & 17.6 &     & DC 5  \\
Lan 188 & 04 16 02.5 & +59 44 02.4 &   6 & $-$22 & 17.5 & $-$0.6 & 17.8 & 18.2 & 18.5 &     & DA 2.0\\
Lan 191 & 04 04 23.8 & +58 58 36.2 & $-$14 &   8 & 16.8 & $-$0.3 & 17.6 & 17.5 & 17.8 &     & DA 5.0\\
Lan 197 & 00 53 01.7 & +59 59 42.6 & $-$76 &   8 & 15.0 & $-$0.6 & 16.4 & 16.2 & 15.8 &     & DA 1.5\\
Lan 253 & 00 33 40.7 & +55 51 47.0 &  14 & $-$14 & 14.8 & $-$0.6 & 16.6 & 16.5 & 15.6 &     & DA 2.5\\
 Lan 272 & 02 57 45.1 & +60 34 27.4 &  10 & $-$12 & 14.0 & $-$0.5 & 15.4 & 15.4 & 15.9 & PN\tablenotemark{f}    & DAO 0.5\\
Lan 273 & 03 08 18.4 & +60 35 30.1 &  14 & $-$74 & 17.0 & $-$0.6 & 17.6 & 17.3 & 17.2 &     & DA 4.0\\
Lan 280 & 02 48 41.3 & +59 16 12.4 &  44 &-118 & 16.5 & $-$0.7 & 16.3 & 16.1 & 16.6 &     & DA 2.5\\
Lan 284 & 03 04 17.9 & +58 44 05.2 & $-$36 & $-$22 & 17.0 & $-$1.2 & 16.5 & 16.6 & 17.0 &     & DA 2.5\\
Lan 351 & 04 06 07.1 & +54 31 33.9 & $-$90 &  16 & 14.5 & $-$0.6 & 15.6 & 15.5 & 15.5 & DA3\tablenotemark{a} & DA 3.0\\
Lan 356 & 20 24 33.0 & +41 23 37.2 &  16 &  12 & 17.0 & $-$0.4 & 17.3 & 17.1 & 17.6 &     & DA 2.5\\
Lan 361 & 20 48 08.3 & +39 51 38.3 & $-$36 & $-$34 & 14.0 & $-$0.9 & 14.8 & 14.5 & 14.2 & DA\tablenotemark{a}  & DA 1.0\\
Lan 364 & 20 56 37.2 & +43 13 26.3 &   2 &  10 & 17.0 & $-$0.2 & 17.8 & 16.5 & 15.9 &     & DZ    \\
Lan 367 & 21 04 02.6 & +42 47 03.5 &  32 &  22 & 15.5 & $-$0.3 & 16.8 & 16.7 & 16.7 & B0-B3\tablenotemark{c} & DA 2.0\\
Lan 401 & 21 16 59.3 & +60 40 30.3 &  18 &  38 & 11.6 & $-$0.1 & 13.7 & 12.8 & 12.0 & Late F-G\tablenotemark{c} & F5 V  \\
Lan 403 & 21 33 14.4 & +60 48 30.4 &  24 &  16 & 17.3 & $-$0.6 & 17.7 & 17.7 & 17:  &     & DA 2.5\\
Lan 406 & 21 24 44.9 & +60 35 17.5 &   8 &  12 & 12.4 & $-$0.1 & 14.2 & 13.0 & 11.9 &     & F6 V  \\
Lan 423 & 21 42 14.7 & +58 54 01.7 & $-$14 &  $-$2 & 16.5 & $-$0.4 & 16.5 & 15.3 & 14.2 & Early-B\tablenotemark{c}    & sdF5  \\
Lan 432 & 21 58 32.5 & +58 04 33.4 &  $-$6 & $-$70 & 18.5 & $-$0.6 & 17.4 & 17.2 & 16.9 &     & sdB   \\
Lan 459 & 22 05 38.2 & +62 24 35.7 & $-$38 &  10 & 16.0 & $-$0.8 & 17.1 & 17.5 & 17.7 & DA\tablenotemark{a,c}  & DA 1.5\\
Lan 480 & 03 29 05.7 & +64 04 42.9 & 29.6&-11.2& 11.3 & $-$0.9 & 13.2 & 13.2 & 13.2 &sdO\tablenotemark{e}  & sdOB  \\
Lan 486 & 03 24 31.3 & +62 50 53.8 & $-$24 &   0 & 15.3 & $-$0.7 & 15.8 & 16.1 & 16.5 &     & DA 3.0\\
Lan 489 & 03 00 37.2 & +62 28 27.9 &   6 &  30 & 16.8 & $-$0.5 & 16.3 & 16.6 & 16.9 &     & DC 5  \\
Lan 532 & 04 12 42.9 & +40 41 26.5 &  14 & $-$44 & 16.0 & $-$0.7 & 16.4 & 16.9 & 17.0 &     & DA 1.5\\
Lan 572 & 04 44 50.1 & +39 15 19.7 & $-$26 & $-$44 & 17.0 & $-$0.5 & 17.1 & 17.5 & 17.5 &     & DA 3.5
 \enddata
\tablenotetext{a}{\citet{MS99}.}
\tablenotetext{b}{\citet{LS77}.}
\tablenotetext{c}{\citet{Era2002}.}
\tablenotetext{d}{\citet{Walker1981}.}
\tablenotetext{e}{\citet{Kilkenny1988}.}
\tablenotetext{f}{\citet{Chromey1979}.}
\end{deluxetable}
\end{landscape}

\begin{deluxetable}{lrrcrrccrr}
\tabletypesize{\scriptsize}
\tablecolumns{10}
\tablewidth{0pt}
\tablecaption{Atmospheric Parameters of DA Stars}
\tablehead{
\colhead{Name} &
\colhead{$T_e$} &
\colhead{(K)} &
\colhead{log $g$} &
\colhead{$M_\odot$} &
\colhead{$M_V$} &
\colhead{log $L_\odot$} &
\colhead{$V_{\rm phot}$} &
\colhead{$D ({\rm pc})$} &
\colhead{log $\tau$} }
\startdata
Lan 004      & 28640    &(  422)   &7.33 (0.05)    &0.38 (0.01)    & 9.00          &$-$0.53        & 16.08&  261&7.04           \\
Lan 018      & 30350    &(  443)   &7.96 (0.05)    &0.63 (0.03)    & 9.87          &$-$0.84        & 12.99&   42&7.02           \\
Lan 019      & 77680    &( 2732)   &7.77 (0.13)    &0.66 (0.04)    & 8.26          &$+$1.00        & 15.54&  285&5.77           \\
 Lan 021      & 55570    &( 1387)   &7.64 (0.09)    &0.56 (0.03)    & 8.41          &$+$0.48        & 15.77&  296&6.19           \\
Lan 022      & 14350    &(  247)   &7.95 (0.05)    &0.58 (0.03)    &11.29          &$-$2.17        & 15.02&   55&8.32           \\
Lan 024      & 12180    &(  317)   &8.26 (0.09)    &0.77 (0.06)    &12.03          &$-$2.64        & 17.09&  103&8.72           \\
Lan 029      & 26780    &(  518)   &8.40 (0.07)    &0.88 (0.05)    &10.86          &$-$1.35        & 17.49&  212&7.76           \\
Lan 062      & 17330    &(  308)   &8.09 (0.05)    &0.67 (0.03)    &11.16          &$-$1.92        & 16.89&  140&8.16           \\
Lan 082      & 14800    &(  387)   &8.46 (0.05)    &0.90 (0.03)    &12.01          &$-$2.43        & 15.38&   47&8.63           \\
Lan 126      & 12380    &(  315)   &8.22 (0.09)    &0.74 (0.05)    &11.93          &$-$2.58        & 17.03&  104&8.67           \\
Lan 159      &  9050    &(  179)   &8.32 (0.17)    &0.80 (0.11)    &13.08          &$-$3.20        & 17.26&   68&9.11           \\
Lan 161      & 16600    &(  985)   &8.39 (0.16)    &0.86 (0.10)    &11.70          &$-$2.18        & 17.52&  145&8.45           \\
Lan 163      & 31810    &(  630)   &7.25 (0.11)    &0.37 (0.03)    & 8.61          &$-$0.28        & 17.23&  530&6.46           \\
Lan 188      & 25160    &( 1542)   &8.27 (0.23)    &0.80 (0.14)    &10.78          &$-$1.38        & 17.98&  276&7.71           \\
Lan 191      &  9870    &(  194)   &8.43 (0.14)    &0.87 (0.09)    &12.93          &$-$3.12        & 17.55&   84&9.10           \\
Lan 197      & 32290    &(  489)   &7.88 (0.06)    &0.59 (0.03)    & 9.61          &$-$0.67        & 16.31&  218&6.88           \\
Lan 253      & 20260    &(  334)   &7.95 (0.05)    &0.60 (0.03)    &10.68          &$-$1.56        & 16.55&  149&7.76           \\
 Lan 272\tablenotemark{a} & 90410 &( 2757) &6.99 (0.13) &0.51 (0.03) & 6.30         &$+$1.93        & 15.40&  659&4.44           \\
Lan 273      & 12960    &(  442)   &7.93 (0.11)    &0.57 (0.06)    &11.45          &$-$2.33        & 17.46&  159&8.44           \\
Lan 280      & 19020    &(  379)   &7.73 (0.06)    &0.48 (0.03)    &10.46          &$-$1.53        & 16.21&  140&7.72           \\
Lan 284      & 20280    &(  532)   &8.12 (0.08)    &0.70 (0.05)    &10.94          &$-$1.66        & 16.55&  132&7.94           \\
Lan 351      & 16750    &(  261)   &8.08 (0.05)    &0.66 (0.03)    &11.19          &$-$1.97        & 15.55&   74&8.20           \\
Lan 356      & 24590    &(  486)   &8.31 (0.06)    &0.82 (0.04)    &10.88          &$-$1.44        & 17.21&  184&7.80           \\
Lan 361      & 66320    &( 1396)   &7.98 (0.07)    &0.72 (0.03)    & 8.85          &$+$0.55        & 14.66&  145&6.00           \\
Lan 367      & 25230    &(  461)   &8.17 (0.06)    &0.73 (0.04)    &10.61          &$-$1.30        & 16.75&  169&7.55           \\
Lan 403      & 19250    &(  450)   &8.39 (0.07)    &0.86 (0.05)    &11.45          &$-$1.92        & 17.70&  178&8.26           \\
Lan 459      & 29650    &(  704)   &7.67 (0.12)    &0.49 (0.05)    & 9.48          &$-$0.70        & 17.28&  364&6.98           \\
Lan 486      & 17600    &(  396)   &8.11 (0.07)    &0.68 (0.04)    &11.16          &$-$1.90        & 15.94&   90&8.15           \\
Lan 532      & 34040    &(  536)   &8.62 (0.06)    &1.02 (0.04)    &10.78          &$-$1.09        & 16.63&  148&7.68           \\
Lan 572      & 14050    &(  817)   &8.22 (0.09)    &0.75 (0.06)    &11.72          &$-$2.37        & 17.28&  129&8.53           
\enddata
\tablenotetext{a}{This object is a DAO star with $\log {\rm He/H}=-1.4$.}
\end{deluxetable}

\begin{deluxetable}{lrrccrrccrr}
\tabletypesize{\scriptsize}
\tablecolumns{10}
\tablewidth{0pt}
\tablecaption{Atmospheric Parameters of DB and DBA Stars}
\tablehead{
\colhead{Name} &
\colhead{$T_e$} &
\colhead{(K)} &
\colhead{log $g$} &
\colhead{log H/He} &
\colhead{$M_\odot$} &
\colhead{$M_V$} &
\colhead{log $L_\odot$} &
\colhead{$V_{\rm phot}$} &
\colhead{$D ({\rm pc})$} &
\colhead{log $\tau$} }
\startdata
Lan 142      & 18020    &(  507)   &8.23 (0.16)    &$-$5.08 (0.37)\tablenotemark{a}       &0.74 (0.10)    &11.24          &$-$1.95        & 17.78&  203&8.23           \\
Lan 143      & 15240    &(  285)   &8.66 (0.11)    &$-$3.71 (0.04)      &1.01 (0.06)    &12.36          &$-$2.53        & 16.80&   77&8.75           \\
Lan 150      & 21000    &(  855)   &7.97 (0.09)    &$-$6.50 (0.00)\tablenotemark{a}       &0.59 (0.05)    &10.57          &$-$1.52        & 17.43&  235&7.76           \\
Lan 158      & 21930    &( 1309)   &7.92 (0.10)    &$-$4.32 (0.36)\tablenotemark{a}       &0.56 (0.05)    &10.44          &$-$1.41        & 16.60&  170&7.61           \\
Lan 174      & 22400    &( 1987)   &7.33 (0.23)    &$-$4.59 (0.86)\tablenotemark{a}       &0.33 (0.06)    & 9.61          &$-$1.02        & 17.15&  321&7.56           
\enddata
\tablenotetext{a}{Upper limit based on the absence of H$\alpha$.}
\end{deluxetable}

\end{document}